\documentclass[%
 reprint,
superscriptaddress,
floatfix,
 amsmath,amssymb,
 aps, prx
]{revtex4-2}

\usepackage[utf8]{inputenc}
\usepackage{amsmath}
\usepackage{amssymb}
\usepackage{graphicx}
\usepackage{empheq}
\usepackage[colorlinks=true, allcolors=blue]{hyperref}

\newcommand*{\vrr}{\mathbf{r}}
\newcommand*{\vp}{\mathbf{p}}
\newcommand*{\vG}{\mathbf{G}}
\newcommand*{\vv}{\mathbf{v}}
\newcommand*{\vu}{\mathbf{u}}
\newcommand*{\SNR}{{\rm SNR}}

\newcommand{\hl}[1]{{\color{red}#1}}

\newcommand{\citeasnoun}[1]{Ref.~\cite{#1}}

\usepackage[english]{babel}
\makeatletter
\def\bbl@set@language#1{%
  \edef\languagename{%
    \ifnum\escapechar=\expandafter`\string#1\@empty
    \else\string#1\@empty\fi}%
  \@ifundefined{babel@language@alias@\languagename}{}{%
    \edef\languagename{\@nameuse{babel@language@alias@\languagename}}%
  }%
  \select@language{\languagename}%
  \expandafter\ifx\csname date\languagename\endcsname\relax\else
    \if@filesw
      \protected@write\@auxout{}{\string\select@language{\languagename}}%
      \bbl@for\bbl@tempa\BabelContentsFiles{%
        \addtocontents{\bbl@tempa}{\xstring\select@language{\languagename}}}%
      \bbl@usehooks{write}{}%
    \fi
  \fi}
\newcommand{\DeclareLanguageAlias}[2]{%
  \global\@namedef{babel@language@alias@#1}{#2}%
}
\makeatother

\DeclareLanguageAlias{en}{english}
\DeclareLanguageAlias{EN}{english}

\begin{document}

\preprint{APS/123-QED}

\title{Bounds on the Coupling Strengths of Communication Channels and Their Information Capacities}

\author{Zeyu Kuang}
\affiliation{Department of Applied Physics and Energy Sciences Institute, Yale University, New Haven, Connecticut 06511, USA}
\author{David A. B. Miller}
\affiliation{Ginzton Laboratory, Stanford University, 348 Via Pueblo Mall, Stanford, California 94305-4088,
USA}
\author{Owen D. Miller}
\affiliation{Department of Applied Physics and Energy Sciences Institute, Yale University, New Haven, Connecticut 06511, USA}

\date{\today}
\begin{abstract}
The concept of optimal communication channels shapes our understanding of wave-based communication. 
Its analysis, however, always pertains to specific communication-domain geometries, without a general theory of scaling laws or fundamental limits.
In this article, we derive shape-independent bounds on the coupling strengths and information capacities of optimal communication channels for any two domains that can be separated by a spherical surface. Previous computational experiments have always observed rapid, exponential decay of coupling strengths, but our bounds predict a much slower, \emph{sub}-exponential optimal decay, and specific source/receiver distributions that can achieve such performance. Our bounds show that domain sizes and configurations, and \emph{not} domain shapes, are the keys to maximizing the number of non-trivial communication channels and total information capacities. Applicable to general wireless and optical communication systems, our bounds reveal fundamental limits to what is possible through engineering the communication domains of electromagnetic waves.
\end{abstract}

\pacs{Valid PACS appear here}
\maketitle

\section{Introduction}
Optimal communication channels define the optimal set of sources and measurements for communicating between two volumes~\cite{miller1998spatial, miller2000communicating, piestun2000electromagnetic, miller2019waves}. The total number and relative strengths of the communication channels depend sensitively on the the size and shape of the volumes, which has restricted most studies to specific and often highly symmetric geometries~\cite{solimene2015singular, leone2018inverse, migliore2006role,leone2019comparison,leone2020inverse,pierri1998information,solimene2007number, solimene2013singular,miller2019waves,solimene2013role, solimene2014inverse,poon2005degrees,leone2018application, leone2019radiation, suryadharma2017singular,piestun2009fundamental,chiurtu2000varying,chiurtu2001capacity,wallace2002intrinsic,jensen2004review,jensen2008capacity,gustafsson2004spectral,glazunov2010physical, ehrenborg2021capacity,ehrenborg2017fundamental,ehrenborg2020physical}, with only overall sum rules known rigorously for arbitrary shapes~\cite{miller2000communicating, miller2019waves}. In this article, we propose a bound on the individual coupling strengths between two domains of \textit{any} shape, as long as a spherical surface separates them. Our theory leverages a monotonicity property of the singular values of the Green's-function operator to bound the strength of each channel by its analytical counterpart from a concentric bounding volume. A key hypothesis about communication channels has been a seemingly universal exponential decay of the coupling strengths with the channel number, supported in numerous computations~\cite{slepian1961prolate, boyd1961confocal,frieden1971viii,bertero1982resolution,solimene2015singular,leone2018inverse, migliore2006role,leone2019comparison, leone2020inverse,pierri1998information, solimene2007number, solimene2013singular,miller2019waves}, and which we rigorously prove in two dimensions. Surprisingly, however, we find that such behavior is \emph{not} generic: in three dimensions our bounds decay sub-exponentially, such that their logarithms decrease only with the square root of the channel number and not linearly. The origin of this decay is the additional degeneracies that are possible for concentric domain configurations in three-dimensional space, which underscores the role of dimensionality in channel counting. Our approach leads directly to shape-independent bounds on two fundamental metrics in communication science: the maximal number of non-trivial channels and their information capacities. The bounds show that increasing domain size and optimizing the global configuration, rather than altering the local patterning of the domain shape, are the keys to increasing the number of non-trivial channels and maximizing their information capacities.

Optimal communication channels represent a unifying framework for optical physics~\cite{miller2019waves, fox1961resonant, miller2007fundamental, miller2007fundamental2, miller2017universal} with a wide range of applications in communication sciences~\cite{telatar1999capacity,goldsmith2003capacity, tse2005fundamentals,vellekoop2007focusing,martinsson2008communication,popoff2010measuring,miller2013establishing, miller2013self,li2014space, zhao2015capacity, miller2017better, annoni2017unscrambling,yilmaz2019transverse}. The Green's-function operator that connects a source volume to a receiver volume, while accounting for all possible background scattering, unambigously identifies the optimal channel profiles and their coupling strengths through its singular vectors and singular values, respectively~\cite{miller1998spatial, miller2000communicating, piestun2000electromagnetic,miller2019waves}. Yet identifying the singular-value decomposition is generically an expensive and opaque computation, which has often limited previous work to highly symmetric domains, with little understanding of general properties or scaling laws~\cite{solimene2015singular, leone2018inverse, migliore2006role,leone2019comparison,leone2020inverse,pierri1998information,solimene2007number, solimene2013singular,miller2019waves,solimene2013role, solimene2014inverse,poon2005degrees,leone2018application, leone2019radiation, suryadharma2017singular,slepian1961prolate, boyd1961confocal, frieden1971viii,bertero1982resolution} other than overall sum rules~\cite{miller1998spatial, miller2000communicating, piestun2000electromagnetic, miller2019waves}.  

A classical example that is analytically solvable is the communication between two identical rectangular or circular apertures in the paraxial limit, where the optimal communicating channels are prolate spheroidal waves, exhibiting exponentially decaying coupling strengths~\cite{slepian1961prolate, boyd1961confocal, frieden1971viii,bertero1982resolution}. 
Similarly rapid decays of channel strengths are observed across different systems, ranging from simple geometries such as rectangular prisms~\cite{miller2000communicating, piestun2000electromagnetic}, strip objects~\cite{solimene2013role, solimene2014inverse,solimene2015singular}, and concentric circumferences~\cite{poon2005degrees,leone2018application, leone2018inverse}, to complex geometries involving conformal conic arcs~\cite{migliore2006role, leone2019radiation,leone2019comparison, leone2020inverse} and multiple rectilinear or spherical domains~\cite{pierri1998information, solimene2007number, solimene2013singular,suryadharma2017singular}. Many of these geometries are reexamined in a recent review paper~\cite{miller2019waves}, where numerical observation of apparent exponential decay of coupling strengths past heuristic limits is hypothesized as being possibly universal.


In addition to the channel-strength decay rate, a related open question has been the maximum total number of channels that can be supported between two regions. Identifying bounds on the number of channels has been of interest since the birth of the field~\cite{miller1998spatial, miller2000communicating, piestun2000electromagnetic, miller2019waves}, with partial success: channel sum rules imply upper bounds on the number of ``well-coupled'' channels simply by assumption of a minimum power-measurement threshold and equal division of power among all channels. Yet, as illustrated numerically, for example, in~\citeasnoun{miller2019waves}, once we move beyond some simple geometries, such as parallel plane surfaces in a paraxial limit, even well-coupled channels can show substantially different power coupling strengths.

An inspiring precursor to our work is that of \citeasnoun{piestun2009fundamental}. In ~\citeasnoun{piestun2009fundamental}, the authors examine the number of communication channels in two dimensions and derive a bound on the number of communication channels between two domains. There is a subtle mathematical issue regarding domain monotonicity (or the lack thereof) of their suggested channel normalization which means that their result is in fact not a fundamental limit (discussed more in the SM), but their result can be understood as a heuristic that identifies the correct scaling laws for circular domains in 2D, and roughly maps to the ultimate fundamental limits. The key insight of ~\citeasnoun{piestun2009fundamental}—Green’s-function singular values have monotonicity properties that imply fundamental limits—forms the foundation of our approach, and enables both the fundamental limits and asymptotic analysis that we identify in two and three dimensions.


The information capacities of optimal communication channels have been investigated for domains of various shapes including spherical~\cite{poon2005degrees}, cubic~\cite{hanlen2006wireless, lee2016capacity}, and non-symmetrical geometries~\cite{tse2005fundamentals, migliore2006role}.
There are shape-dependent bounds to the information capacities for line-of-sight communications~\cite{chiurtu2000varying, chiurtu2001capacity,wallace2002intrinsic, jensen2004review,jensen2008capacity} and spherical communication domains~\cite{gustafsson2004spectral,glazunov2010physical}.  
A more general computational framework is proposed in Refs.~\cite{ehrenborg2017fundamental,ehrenborg2020physical,ehrenborg2021capacity} which bounds the total information capacity of a communicating domain by optimizing over freely varying currents in a larger bounding domain.
This approach, however, forcefully restricts the optimization space of the free-varying currents to bound the information capacity of a MIMO system that only has access to $N$ number of communication channels~\cite{ehrenborg2020physical}.
Such restrictions yield an optimization space, represented by the span of the first $N$ optimal communication channels of the bounding domain, that  does not encompass all possible current distributions of any $N$-channel MIMO system.

Thus the questions remain: how rapidly must optimal-communication-channel strengths decay, what is the maximum number of usable communication channels, and does this imply bounds on maximum information capacities? We answer each of these questions below.

\section{Optimal communication channels}
\label{sec:communication_channel}
Our shape-independent bounds stem from combining two known theorems: (1) the coupling strengths of the optimal communication channels between two domains are the singular values of the corresponding Green's-function operator~\cite{miller1998spatial, miller2000communicating, piestun2000electromagnetic, miller2019waves} and (2) those singular values increase monotonically with the size of the two domains~\cite{hanson2013operator}. The singular value decomposition of the dyadic Green's-function operator $\vG(\vrr,\vrr')$ from a source region $V_s$ to a receiver region $V_r$ is
\begin{equation}
    \vG(\vrr,\vrr') = \sum_{q=1}^{\infty} s_q \vu_q(\vrr)\vv_q^*(\vrr'),
  \label{eq:SVD_G}
\end{equation}
where $\{\vv_q(\vrr)\}_{q=1}^{\infty}$ is a set of orthonormal vector-valued basis functions in the source region, $\{\vu_q(\vrr)\}_{q=1}^{\infty}$ is a set of orthonormal vector-valued basis functions in the receiver region, and $\{s_q\}_{q=1}^{\infty}$ is the set of (non-negative) singular values. The tuples $\{(\vv_q,\vu_q,s_q)\}_{q=1}^{\infty}$ represent the optimal communication channels, with the fields radiated from sources $\vv_q(\vrr)$ mapping uniquely to fields $\vu_q(\vrr)$ in the receiver region with amplitudes $s_q$. The absolute square of the amplitude $|s_q|^2$ is referred to as the coupling strength or channel strength of channel $q$.

The key theorem that enables our shape-independent bounds, and is perhaps less well-known, is that all singular values of a Green's-function operator, as in Eq.~(\ref{eq:SVD_G}), may not decrease as the source and receiver domains are enlarged~\cite{hanson2013operator}. More precisely: if one domain encloses another, each singular value of the former cannot be smaller than the corresponding singular value of the latter.  We refer to this property of coupling strengths as ``domain monotonicity.'' It can be proven through a recursive argument. To simplify the notation and intuition, we encode the spatial variations of the source amplitude and polarization in a finite-dimensional vector $\vp$, define the Green's-function operator as a finite-dimensional matrix $\vG$, and use $\dagger$ to denote Hermitian conjugation. The operators $\vG^\dagger \vG$ and $\vG\vG^\dagger$ are necessarily Hermitian operators, which means their eigenvalues are real, and their eigenvalue/eigenfunction pairs can be found variationally via maximization and orthogonalization. (Another important characteristic of these operators is that they are positive semidefinite which implies their eigenvalues are non-negative, though this positive-semidefinite property is not a necessary condition for the variational procedure we discuss below.)
The square of the first singular value is obtained by maximizing the Rayleigh quotient of $\vG^\dagger\vG$: $|s_1|^2 = \max \frac{\vp^\dagger\vG^\dagger\vG\vp}{\vp^\dagger\vp}$. Clearly this may not decrease as the source domain enlarges, as maximization over a larger space of vectors cannot lead to a smaller optimal value. The second singular value similarly maximizes the Rayleigh quotient, now subject to orthogonality to the first singular vector. Because the first singular vector has changed with the domain, there is not a straightforward comparison to the optimization problem defining the second singular vector of the original domain. Yet the extra freedom given to the first singular vector ultimately only reduces the effect of the orthogonality constraint, such that the second singular value must also increase due to the domain enlargement. (A more precise version of this argument is given in Ref. \cite{molesky2020fundamental}.) The same argument recursively applies to the rest of the singular values, and also for an enlarged receiver domain. Hence we have the key theoretical ingredients: optimal communication channels are defined by the singular-value decomposition of the Green's-function operator between source and receiver domains, and the singular values satisfy domain monotonicity on both domains.

\begin{figure}[t!]
\centering
\includegraphics[width=0.4\textwidth]{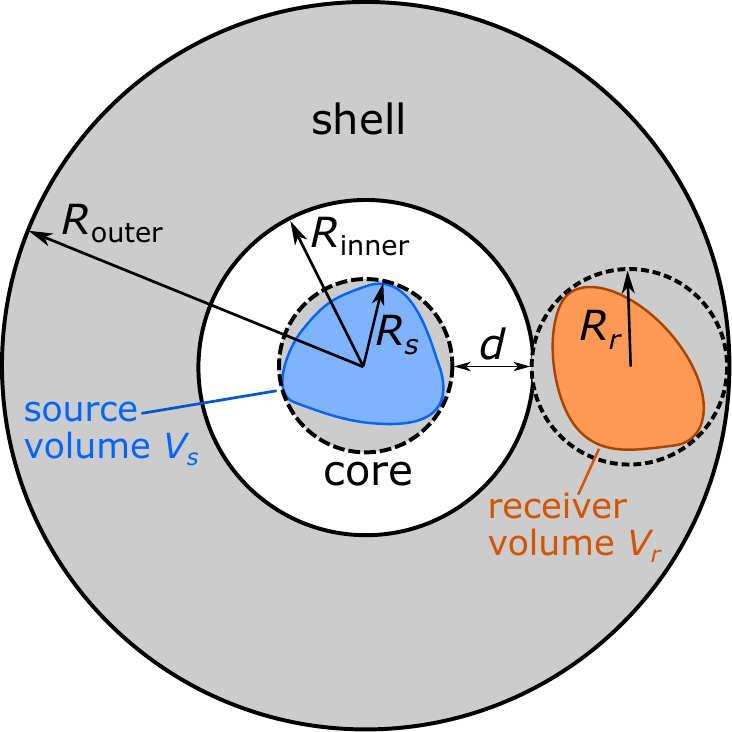}
\caption{The coupling strengths of the communication channels between a source volume $V_s$ and a receiver volume $V_r$ are upper bounded by their counterparts from the core--shell bounding volume (shaded in grey). We can also interchange the roles of the source and receiver volumes, and we obtain tighter bounds by using whichever is smaller as the ``inner'' volume in this figure.}
\label{fig:BoundingVolume}
\end{figure}

\section{Channel-strength bounds}
\label{sec:channel_strength}
In this section, we derive shape-independent bounds on total channel strengths, relative channel strengths normalized against a sum rule, and their collective asymptotic decay rates in the many-channel limit. The domain-monotonicity principle discussed above immediately leads to bounds: the coupling strengths $|s_q|^2$ for arbitrary source and receiver domains are individually bounded above by the respective coupling strengths of any enclosing domains. We select an analytically tractable core--shell set of enclosing domains, depicted as the grey shaded region in Fig.~\ref{fig:BoundingVolume}, which yield the bounds:
\begin{equation}
    |s_{q}|^2 \leq |s_q^{\text{(core--shell)}}|^2,\quad  \text{for}\ q = 1, 2, ...
    \label{eq:sl_max}
\end{equation} 
In such core--shell configurations we can choose either the source or the receiver to be enclosed in the core; to find the tightest upper bounds, we take the minimum of both possible configurations. The core is a cylinder for 2D and a sphere for 3D. In the following sub-sections, we derive analytical expressions for the bounds in both dimensions.

\begin{figure*}[htb]
\centering
\includegraphics[width=1\textwidth]{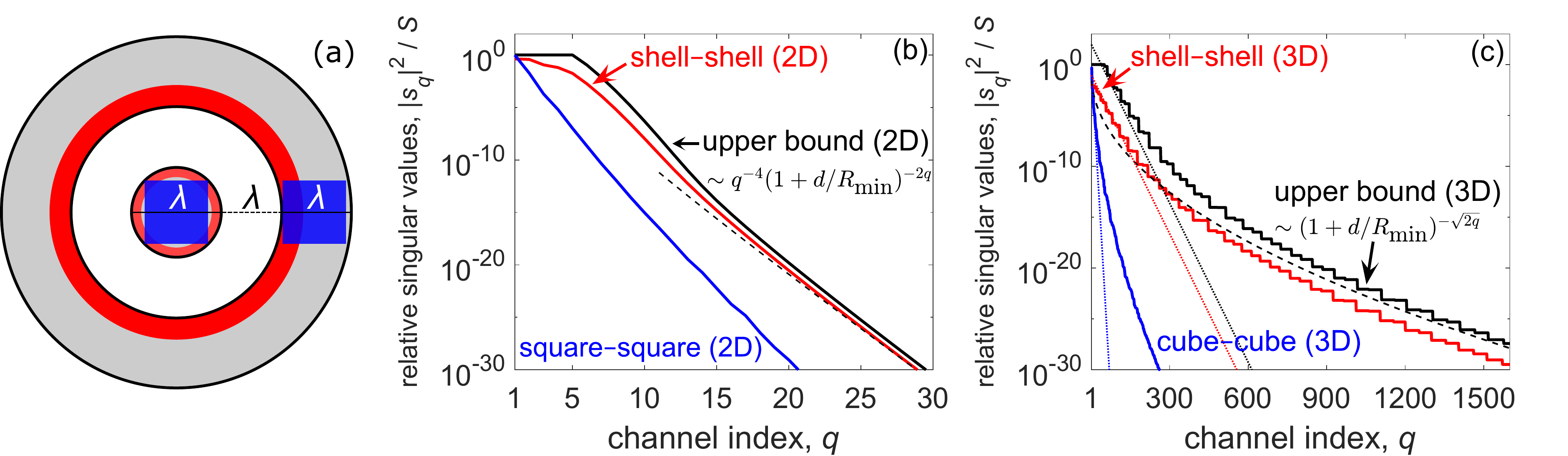}
\caption{\label{fig:sv_bounds} Shape-independent upper bounds on the relative coupling strengths $|s_q|^2$ normalized against the total sum rule $S$ in two- and three-dimensional spaces. (a) A grey-shaded concentric core--shell bounding volume enclosing a square--square configuration of sources and receivers in the blue shaded region, as well as a shell--shell configuration in the red shaded region. (b) In two-dimensional (2D) space, the upper bound, calculated for the grey-shaded bounding volume in (a), decays exponentially as in the dashed black line. (c) In three-dimensional (3D) space, the additional azimuthal degeneracy leads to an optimal sub-exponential decay that is achieved by the shell--shell configuration. The sub-exponential decay rate (dashed black line) suggests many more communication channels at the large-channel limit than previously hypothesized exponential decays (dotted lines).}

\label{fig:Fig2}
\end{figure*}

\subsection{Channel-strength bounds in 2D}
Consider communication in two dimensions between a source domain $V_s$ and a receiver domain $V_r$ as in Fig.~\ref{fig:BoundingVolume}.  The sources are bounded within a cylindrical core of radius $R_s$ and the receivers sit a minimum distance $d$ and maximum distance $d_{\rm max} = d+2R_r+2R_r$ from the sources. The bounding volumes, comprising an inner cylinder and an outer shell, are shaded in grey in Fig.~\ref{fig:BoundingVolume}. 
The singular values of the Green's function operator between the concentric cylinder--shell bounding volume can be identified by first performing a separation of variables for the two-dimensional scalar Green's function $G(\mathbf{r},\mathbf{r}') = \frac{ik^2}{4} H^{(1)}_0(k|\mathbf{r}-\mathbf{r}'|)$ in polar coordinates~\cite{abramowitz1964handbook}:
\begin{equation}
    G(\mathbf{r},\mathbf{r}') = \frac{ik^2}{4} \sum_{q=-\infty}^{\infty} H_q^{(1)}(k\rho)e^{-iq\phi}J_q(k\rho')e^{iq\phi'},
    \label{eq:G2D}
\end{equation}
where the functions $H_q^{(1)}(k\rho)e^{-iq\phi}$ and $J_q(k\rho)e^{-iq\phi}$ are the outgoing and regular cylindrical waves, with $H_q^{(1)}(x)$ and $J_q(x)$ being the Hankel function of the first kind and the Bessel function, respectively. Their polar coordinates $(\rho, \phi)$ and $(\rho', \phi')$ are defined on the bounding shell and bounding cylinder, respectively, relative to the center of the cylinder-shell bounding volume. The cylindrical waves $H_q^{(1)}(k\rho)e^{-iq\phi}$ and $J_q(k\rho)e^{-iq\phi}$ are the (unnormalized) left and right singular vectors of the Green's function operator in the cylinder-shell bounding volume.  (The cylindrical symmetry of the bounding volume ensures orthogonality.) 
There are two possible cylinder--shell bounding volumes: one centers around the source domain and one centers around the receiver domain. To tighten the upper bound, we choose the smaller of the two domains as the "inner" volume in Fig.~\ref{fig:BoundingVolume} because it leads to a smaller coupling strength $|s_{q}^{\text{(cylinder--shell)}}|^2$ which is the product of the norms of the unnormalized singular vectors, $H_q^{(1)}(k\rho)e^{-iq\phi}$ and $J_q(k\rho)e^{-iq\phi}$, in their respective bounding volumes:
\begin{equation}
    |s_{q}^{\text{(cylinder--shell)}}|^2 = \pi^2 k^2  \int_0^{R_{\text{min}}} |J_q(k\rho)|^2 \rho d\rho
    \int_{R_{\rm inner}}^{R_{\rm outer}} |H_q^{(1)}(k\rho)|^2 \rho d\rho.
    \label{eq:snm_Rmin2D}
\end{equation}
As the inner bounding cylinder is chosen to encompass the smaller domain, its radius is the smaller of the two radii, i.e., $R_{\text{min}} = {\rm min}\{R_s, R_r\}$. Similarly, one can show that the inner and outer radii of the outer bounding shell are $R_{\text{inner}} = d+R_{\rm min}$ and  $R_{\text{outer}} = d+2R_s+2R_r-R_{\text{min}}$, respectively.
The singular values in Eq.~(\ref{eq:snm_Rmin2D}) are dimensionless quantities because our Green's function, of Eq.~(\ref{eq:G2D}), differs from the conventional definition \cite{jackson1999classical, miller2019waves} by a factor of $k^2$ to be inversely proportional to volume.

The number of non-trivial communication channels is determined by the number of channels whose \textit{relative} channel strengths are above a certain measurement threshold. The relative channel strengths can be normalized either by a total sum rule $S=\sum_{q=-\infty}^\infty|s_q|^2$ or by the largest channel strength~\cite{miller2019waves}. Lower bounds on the sum rule can be analytically derived based on the monotonic decay of wave energy in free space, thus leading to bounds on the total number of channels above a certain sum-rule energy fraction. The sum rule $S$ is a double integral of the absolute square of the two-dimensional Green's function over both the source and receiver domains~\cite{miller2000communicating, miller2019waves}
\begin{equation}
    S = \int_{S_s}\int_{S_r} |G(\vrr,\vrr')|^2 d\vrr d\vrr' 
      \geq k^4S_sS_r|H_0^{(1)}(kd_{\rm max})|^2/16,
      \label{eq:sum2D}
\end{equation}
where we further lower bound $S$ by the fact that the magnitude of the Green's function takes its minimal value at the most separated points between the two domains, which is at a distance $d_{\rm max} = d + 2R_r + 2R_s$ for the cylinder--shell bounding volume illustrated in Fig.~\ref{fig:BoundingVolume}. The variables $S_s$ and $S_r$ in Eq.~(\ref{eq:sum2D}) denote the total area of the source and receiver domains. Combining Eq.~(\ref{eq:snm_Rmin2D}) and Eq.~(\ref{eq:sum2D}), we derive 
\begin{align}
    \frac{|s_{q}|^2}{S} \leq \frac{16|s_{q}^{\text{(cylinder--shell)}}|^2 }{k^4S_sS_r|H_0^{(1)}(kd_{\rm max})|^2},
    \label{eq:sr_2D}
\end{align}
which is a shape-independent bound on the relative channel strength between domains in two-dimensional space. In the many-channel limit, the bound in Eq.~(\ref{eq:sr_2D}) simplifies:
\begin{equation}
    \frac{|s_{q}|^2}{S} \leq \frac{R_{\text{min}}^4}{q^4S_sS_r|H_0^{(1)}(kd_{\rm max})|^2(1+d/R_{\text{min}})^{2(q-1)}},\ \text{as}\ q \rightarrow \infty.
    \label{eq:decay2D}
\end{equation}
The presence of the exponential factor of $2(q-1)$ indicates that channel strengths in two dimensions must decay at least exponentially fast with channel number, in agreement with the previously hypothesized exponential decay of channel strengths. The exponential decay rate depends only on the separation distance $d$ relative to the smaller radius $R_{\text{min}}$ between the two communication domains.

The upper bound in Eq. (\ref{eq:sum2D}) and its optimal exponential decay in Eq. (\ref{eq:decay2D}) applies to any two domains that can be separated by a cylindrical surface. The bound is achieved by concentric communicating domains that fill the bounding volume, while the optimal decay rate can also be achieved with concentric sub-domains. To illustrate the latter point, in Fig. \ref{fig:Fig2}(a), we arrange a fixed number of sources and receivers in two different configurations inside a bounding volume. The first configuration (blue shaded region) consists of two squares of sources and receivers with the side lengths of $\lambda/\sqrt{2}$. The second configuration (red shaded region) consists of concentric shell-like communicating domains with the same source and receiver areas. Both configurations are enclosed in a concentric cylinder--shell bounding volume of $2R_s=2R_r=d=\lambda$. Inside this bounding volume, the maximal relative coupling strength is given by the solid black line in Fig. \ref{fig:Fig2}(b), calculated using Eq. (\ref{eq:sr_2D}).   We observe that, while the square--square configuration (solid blue line) falls far short of the bound, arranging the same number of sources and receivers to cover a wider solid angle in a shell--shell configuration  (solid red line) enables close approach to the upper bound. (The black-line upper bound is clamped to 1; no channel can have strength larger than 1. The looseness of Eq. (\ref{eq:decay2D}) arises from the dramatic mismatch of the source--receiver volumes to the bounding volumes.)  Moreover, the shell--shell configuration achieves the optimal exponential decay predicted in Eq. (\ref{eq:decay2D}). This result corroborates previous works~\cite{slepian1961prolate, boyd1961confocal,frieden1971viii,bertero1982resolution,solimene2015singular,leone2018inverse, migliore2006role,leone2019comparison, leone2020inverse,pierri1998information, solimene2007number, solimene2013singular,miller2019waves} that predicted exponential decay in wide-ranging scenarios, and hypothesized that exponential decay may be a universal rule. As we show below, however, the three-dimensional behavior is quite different.


\subsection{Channel-strength bounds in 3D}
\label{ssc:channel3D}
The derivation of shape-independent bounds on channel strengths in three dimensions is similar to the derivation in two dimensions, with the cylinders replaced by spheres. For this 3D case, we now use a full vector formulation of the problem, as appropriate for a full electromagnetic solution. So, we move to dyadic Green's functions, and we start by expanding the dyadic Green's function as a summation of outer products, now of spherical vector waves: 
\begin{equation}
    \vG(\vrr, \vrr') = ik^3 \sum_{n=0}^\infty \sum_{m=-n}^n \sum_{j=1,2} \vv_{\text{out}, nmj}(\vrr)\vv_{\text{reg}, nmj}^*(\vrr'),
    \label{eq:Gr3D}
\end{equation}
where $\vv_{\text{out}, nmj}$ and $\vv_{\text{reg}, nmj}(\vrr')$ are the outgoing and regular spherical vector waves~\cite{tsang2004scattering} defined on the bounding shell and bounding sphere\hl{,} respectively.
The vectors $\vrr$ and $\vrr'$ are spherical coordinates defined with respect to the center of the concentric bounding volume.
The regular (outgoing) spherical vector waves are formed by combining the angular dependency of vector spherical harmonics with the radial dependency of spherical Bessel (Hankel) functions~\cite{tsang2004scattering}.
Explicit expressions of the vector spherical waves, $\vv_{\text{out}, nmj}$ and $\vv_{\text{reg}, nmj}(\vrr)$, and the wave equation we use to define the Green's function are given in the SM.
The indices $n$ and $m$ index the underlying spherical harmonics, and $j=1,2$ denotes the two possible polarizations of a transverse vector field.
The orthogonality of the spherical waves in a spherically symmetric domain allows us to identify $\vv_{\text{out}, nmj}$ and $\vv_{\text{reg}, nmj}(\vrr)$ as the (unnormalized) left and right singular vectors of the Green's function operator defined on the three-dimensional sphere-shell bounding volumes. 
The corresponding singular values are the products between the norms of functions $\vv_{\text{out}, nmj}$ and $\vv_{\text{reg}, nmj}(\vrr)$ in their respective volumes:
\begin{align}
    |s_{nmj}^{\text{(sphere--shell)}}|^2 = k^6 & \int_{V_{\text{shell}}} \left|\vv_{\text{out}, nmj}(\vrr)\right|^2 d\vrr \nonumber \\ &\cdot\int_{V_{\text{sphere}}}  \left|\vv_{\text{reg}, nmj}(\vrr)\right|^2 d\vrr,
    \label{eq:snm}
\end{align}
where $V_{\text{shell}}$ and $V_{\text{sphere}}$ represent the volumes of the bounding shell and bounding sphere.
Explicit expressions of the singular values $|s_{nmj}^{\text{(sphere--shell)}}|^2$ can be found in the SM. According to the domain-monotonicity property in Eq.~(\ref{eq:sl_max}), the $q$-th largest number from the set of all possible $|s_{nmj}^{\text{(sphere--shell)}}|^2$ upper-bounds the $q$-th largest channel strength of any configuration of sources and receivers in the sphere--shell bounding volume. 

Again, the number of non-trivial communication channels is determined by normalizing the channel strengths to the total sum rule. The sum rule is now lower bounded by (cf. SM):
\begin{align}
    S = \int_{V_s}\int_{V_r} ||\vG(\vrr,\vrr')||_F^2 d\vrr d\vrr' \geq \frac{k^4V_sV_r}{8\pi^2d_{\rm max}^2} + \mathcal{O}\left(\left(kd_{\rm max}\right)^{-4}\right).
      \label{eq:sum}
\end{align}
For conciseness, we assume the furthest separated points are in the far field, i.e. $kd_{\rm max} \gg 1$, so  that only the leading term in Eq.~(\ref{eq:sum}) remains. This can be easily generalized by explicitly including two other higher-order terms, leading to a somewhat more complicated expression but the same asymptotic properties.

By combining the upper bound of channel strengths in Eq.~(\ref{eq:sl_max}) and the lower bound of the sum rule in Eq.~(\ref{eq:sum}), we derive a key result for 3D communication domains, a shape-independent upper bound on their relative channel strengths normalized against the total sum rule:
\begin{equation}
    \frac{|s_{nmj}|^2}{S} \leq \frac{8\pi^2d_{\rm max}^2}{k^4V_sV_r} |s_{nmj}^{\text{(sphere--shell)}}|^2,
    \label{eq:sr}
\end{equation}
where the singular value of the sphere--shell bounding volume, $|s_{nmj}^{\text{(sphere--shell)}}|^2$, is identified in Eq.~(\ref{eq:snm}), and whose explicit expression can be found in the SM.
One immediate prediction of the upper bound in Eq.~(\ref{eq:sr}) is an optimal sub-exponential decay rate of the channel strengths between two 3D domains, which we now derive. The total number of channels that has $n$-index less or equal to $n$ is $q=2(n+1)^2$. We use this total channel index $q$ as our new index for channel strengths to meaningfully describe their decay rate. When the total number  $q\rightarrow\infty$, Eq.~(\ref{eq:sr}) can be simplified to (cf. SM):
\begin{equation}
    \frac{|s_{q}|^2}{S} \leq \frac{2\pi^2d_{\rm max}^2}{k^4V_sV_r(1+d/R_{\text{min}})^{\sqrt{2q}+1}}, \quad \text{as}\ q \rightarrow \infty,
    \label{eq:decay3D}
\end{equation}
where the parameter $R_{\text{min}} = \min\{R_s, R_r\}$ denotes the radius of the smaller domain.
Equation~(\ref{eq:decay3D}) shows that, regardless of the domain shape, channel strengths $|s_q|^2$ in three-dimensional space have to decay at least as fast as $a^{-\sqrt{q}}$, where $a$ is a bounding-domain-dependent numerical constant ($a=(1+d/R_{\rm min})^{\sqrt{2}}$), and the key new feature is the square root dependence on $q$ in the exponent. 
Such a decay is \emph{sub-exponential}, as its logarithm decays only with the square root of the channel number rather than the (much faster) linear reductions characteristic of exponential decay. 

\begin{figure}[t!]
\centering
\includegraphics[width=0.5\textwidth]{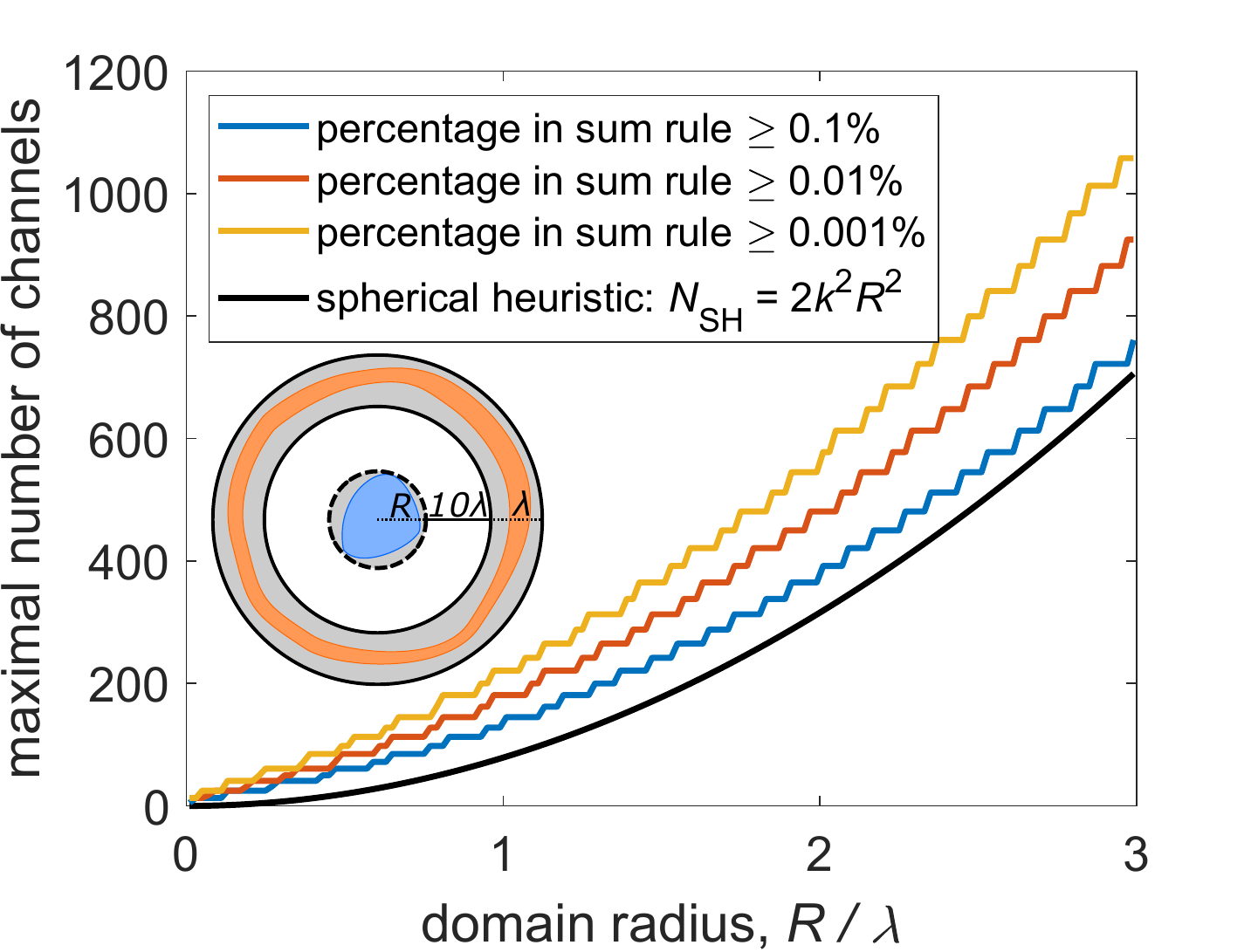}
\caption{ The maximal number of non-trivial communication channels for a domain of maximal radius $R$ and certain measurement thresholds set by their percentage in the total sum rule. The other communication domain is in a bounding shell shown in the inset. 
The quadratic dependence of the bound regards to the domain radius $R$ can be conveniently modelled by a spherical heuristic number, $N_{\text{SH}} = 2k^2R^2$. }
\label{fig:NoC}
\end{figure}

Figure~\ref{fig:Fig2}(c) compares the coupling-strengths bound in 3D, with a clearly sub-exponential decay rate, to the coupling strengths of two configurations of sources and receivers (shell--shell and cube--cube) in a sphere--shell bounding volume. Both configurations possess a volume $\lambda^3 / (3\sqrt{3})$ of sources and receivers and follow the same layout as in Fig.~\ref{fig:Fig2}(a).
Similar to the 2D case, we observe that the shell--shell configuration closely follows the bound while the cube--cube configuration falls short.
Interestingly, both the cube-cube and shell-shell configurations and the upper bound first enter a phase of approximately exponential decay (dotted lines in Fig.~\ref{fig:Fig2}(b), a phenomenon also observed in Ref.~\cite{miller2019waves}) before they exhibit different sub-exponential decays on a larger scale. 
By ``sub-exponential'', we mean that the fall off in the channel strengths is not as fast as exponential; high-index channels have somewhat stronger coupling strength than an exponential fall-off would predict, and, 
in the many-channel limit, the asymptotic sub-exponential decay predicted in Eq.~(\ref{eq:decay3D}) bounds all geometries and also puts forth the concentric shell-shell configurations as the optimal candidate for achieving the slowest sub-exponential decay.

The sub-exponential decay of the channel-strength bound in 3D is in stark contrast with its exponentially decaying counterpart in 2D.
This point is accentuated by contrasting Fig.~\ref{fig:Fig2}(c) with Fig.~\ref{fig:Fig2}(b), where their asymptotic decay rates, shown as the black dashed lines, are fundamentally different. 
This difference originates from the additional azimuthal degeneracy of communication channels in 3D. Such degeneracy manifests through the staircase behavior of the upper bound in Fig.~\ref{fig:Fig2}(c). It allows one to potentially establish many more useful orthogonal channels in 3D: the bound suggests approximately 145 channels for 3D domains above a threshold of $10^{-4}$ in Fig.~\ref{fig:Fig2}(c), as compared to only 8 channels in the 2D case above the same threshold.
The difference in the decay rate of upper bounds in two- and three-dimensional spaces underscores the role of dimensionality in channel counting.

\subsection{Bounds on the number of non-trivial channels}
The number of non-trivial communication channels is often regarded as the number of ``spatial degrees of freedom'' for communicating between two regions, an idea that generalizes the concept of diffraction limits~\cite{miller2019waves} and dictates fundamental response in many wave systems~\cite{miller2007fundamental,miller2015shape, venkataram2020fundamental}. A communication channel is considered non-trivial if its coupling strength is above a certain percentage in the total sum rule~\cite{miller2019waves}; the bounds of Eqs.~(\ref{eq:sr_2D}) and (\ref{eq:sr}) on relative coupling strengths therefore directly lead to bounds on the number of communication channels.

Figure~\ref{fig:NoC} shows the maximal number of channels available for any source domain within a three-dimensional sphere of radius $R$, computed from Eq.~(\ref{eq:sr}). The receiver domain is a shell ten wavelengths away, with a thickness of one wavelength, as shown in the inset of Fig.~\ref{fig:NoC}. (The source and receiver domains can be transposed.) We also assume both domains occupy at least half of their respective bounding volumes. The bounds are plotted as a function of the maximal domain radius $R$ for a number of measurement thresholds. The bounds are not overly sensitive to the measurement threshold: a hundredfold increase in the sensitivity, as occurs going from the blue line to the yellow line, does not even double the number of available channels. On the other hand, the bounds increase approximately quadratically with the maximal domain radius $R$, suggesting enlarging domain size is the key to gaining more useful channels.

The quadratic increase of the bound with respect to the domain radius $R$ can be understood as arising from the increasing surface area of two sufficiently separated communication domains. At first, one might expect the mode number to increase with the \emph{volume} of the domains, but the waves in the volumes are determined by the waves at the surfaces (by the surface equivalence principle~\cite{jin2011theory}), and restrictions on the number of unique wave patterns at the surface will naturally constrain the number of independent volume functions as well. As the domain size increases, we can use the notion of a ``spherical heuristic number,'' denoted $N_{\rm SH}$, to estimate the number of communication channels:
    \begin{equation}
        N_{\rm SH} = 2k^2R^2. \label{eq:SH}
    \end{equation}
Spherical heuristic numbers were proposed in \citeasnoun{miller2019waves}, where the expression $16\pi R^2/\lambda^2$ was suggested. Here we modify the expression, instead assuming one unique spatial mode per $\lambda^2 / \pi$ area on the surface of the spherical bounding domain (instead of $\lambda^2/4$ as previously suggested~\cite{miller2019waves}), multiplied by two polarizations, resulting in the expression of Eq.~(\ref{eq:SH}). The $\lambda^2/\pi$ area expression comes from treating each surface patch on the source and receiver domains as interacting in the paraxial limit--certainly not exactly true, but sufficient to gain intuition. Fig. \ref{fig:NoC} shows quantitative agreement between the spherical heuristic number and the rigorously calculated bound under a $0.1\%$ threshold on the sum rule, explaining the approximately quadratic increase of the number of channels as a function of domain radius.

The bounds in Fig.~\ref{fig:NoC} weakly depend on the sum-rule percentage threshold because of the rapid decay of channel strength at large-order channels. Though not shown in this graph, the bound barely depends on the depth of the receivers and their distance from the source (unless in the extreme near-field limit when the separation distance is much less than a wavelength). All these imply that the group of bounds shown in Fig.~\ref{fig:NoC} represent the intrinsic number of channels one can couple out of any source domain of a given size to the far field.

\section{Bounds on the information capacities of communication channels}
\label{sec:channel_number}
Information capacity, defined as the maximal rate at which the information can be reliably transmitted between two communicating domains, is a notion that has been central to the development of modern communication systems~\cite{shannon1948mathematical,tse2005fundamentals}.
In this section, we show how our coupling-strengths bounds can help one determine the maximal information capacity of communication channels in three-dimensional space.
A key feature of our approach is that it tightly bounds the total information capacity of any given number of channels, which is highly relevant to modern MIMO systems that usually have access to a finite number of antennas.

The information capacity $C$ of $N$ optimal communication channels (per unit time and unit bandwidth) is the sum of the capacity of each channel, each of which logarithmically depends on its input power $P_q$, coupling strength $|s_q|^2$, and noise power $P_{\rm noise}$ \cite{tse2005fundamentals}:
\begin{equation}
    C = \sum_{i=q}^{N}\log_2 \bigg(1+\frac{P_q|s_q|^2}{P_{\rm noise}}\bigg)\quad \text{bits/s/Hz},
    \label{eq:ShannonMulti}
\end{equation}
where we assume an additive white Gaussian noise background with the same noise power $P_{\rm noise}$ for each channel.

A larger domain size is always favorable to increase the information capacity of the first $n$ optimal communication channels. This is because the capacity $C$ in Eq.~(\ref{eq:ShannonMulti}) increases monotonically with coupling strength $|s_q|^2$, which in turn increases monotonically with the domain size. Therefore, the capacity of the sphere--shell bounding volume serves as an upper bound for the capacity of all possible sub-domains within:
 \begin{equation}
    C \leq \sum_{q=1}^{N}\log_2 \bigg(1+\frac{P_q|s_q^{(\text{sphere--shell})}|^2}{P_{\rm noise}}\bigg)\quad \text{bits/s/Hz},
    \label{eq:CapacityBound}
\end{equation}
where the coupling strength $|s_q^{(\text{sphere--shell})}|^2$ of the sphere--shell bounding volume is given in Eq.~(\ref{eq:snm}). One can solve for the optimal allocation of powers $P_q$ for a fixed total power input $\sum_{q=1}^{N} P_q = P$, by the ``water-filling'' algorithm~\cite{tse2005fundamentals}, with the semi-analytical form 
$P_q = \max\{0,\mu - P_{\rm noise}/|s_q^{(\text{sphere--shell})}|^2\}$, where $\mu$ is the numerical constant for which $\sum_{q=1}^{N}P_q = P$. The signal-to-noise ratio (SNR), defined as the ratio between the total power and noise power, i.e. $\SNR = P / P_{\rm noise}$, is the key external parameter that affects the optimal strategy of the power allocation.

\begin{figure}[t!]
\centering
\includegraphics[width=0.5\textwidth]{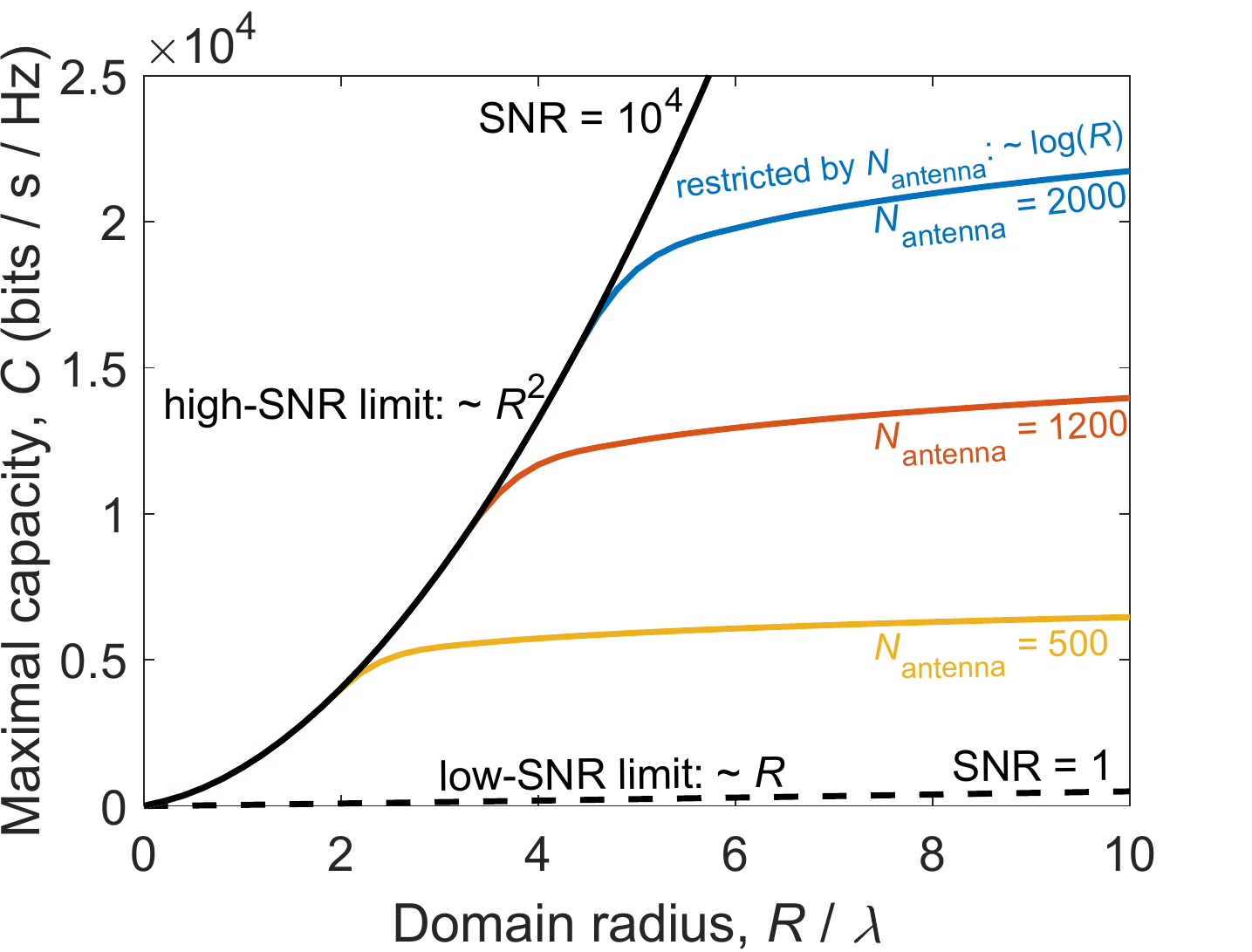}
\caption{\label{fig:CapacityBound} 
Maximal information capacity $C$ between any two domains that fit within the radius-$R$ sphere and wavelength-thickness bounding shell of Fig.~\ref{fig:NoC}. In the high SNR limit, the bound increases quadratically with domain size $R$ (solid black), whereas in the low SNR limit, the bound increases only linearly (dashed black). When the number of available channels is restricted  by the number of antennas, $N_{\rm antenna}$, the channel-capacity bounds tail off and increase only logarithmically with domain size (blue, orange, yellow).} 
\end{figure}

Figure~\ref{fig:CapacityBound} shows the capacity bound for communication between arbitrary domains contained in the sphere--shell bounding volumes in two limits: high SNR (solid black), and low SNR (dashed black). The size dependencies of the capacity bounds are quite different in the two limits. When SNR is very small, the logarithms approximately become linear functions of the power, in which case the optimal allocation puts all of the power in the single channel with the highest coupling strength~\cite{tse2005fundamentals}. The maximum coupling strength scales linearly with the radius $R$:  $\max\left\{|s_{q}^{\text{(sphere--shell)}}|^2\right\} = k^2R_rR$, provided that the radius $R$ of the bounding sphere is much larger than a wavelength and the bounding shell is in the far field of the bounding sphere (cf. SM). Then we have
\begin{equation}
    C \leq \SNR\cdot \log_2(e)  k^2 R_r R,\quad \text{for}\ \ \SNR \rightarrow 0.
\end{equation}
By contrast, in the high-SNR limit, the optimal allocation of power equally divides amongst all channels with nonzero channel strengths~~\cite{tse2005fundamentals}. The information capacity in this case scales with the number of such channels, which, as we established in the Sec.~\ref{sec:channel_number}, depends quadratically with the domain radius $R$ (modeled by the spherical heuristic number $N_{\rm SH} = 2k^2R^2$). Hence, the capacity bound increases quadratically with $R$ in the high-SNR limit:
\begin{equation}
    C \leq 2\log_2({\rm SNR})k^2R^2,\quad \text{for}\ \ \SNR \gg 0.
\end{equation}
In many scenarios, the number of communication channels may be restricted well below our electromagnetic limit; one common example may be a MIMO system with antennas spaced more than half a wavelength apart. When the number of communication channels is restricted by the number of antennas, $N_{\text{antenna}}$, the growth in the large-domain limit cannot remain quadratic or even linear; instead, the capacity bound will grow logarithmically at best. This is because for a fixed number of channels, the capacity of each channel increases logarithmically with channel strength, which in turn increases at most linearly with $R$:
\begin{equation}
    C \leq N_{\rm antenna}\log_2\left(\SNR\cdot\frac{ k^2R_rR}{N_{\rm antenna}}\right),\quad \text{for}\ \ \SNR \gg 0.
\end{equation}
The  logarithmic dependence is confirmed by the computations of the blue, orange, and yellow lines in Fig.~\ref{fig:CapacityBound}, with each having the same SNR as the solid black line, but decreasing $N_{\rm antenna}$. The quadratic increase at the outset of each curve saturates almost exactly at the domain size where the number of electromagnetic channels ($N_{\rm SH} = 2k^2 R^2$) equals $N_{\rm antenna}$. Thus, despite the abundant number of electromagnetic channels in a large domain, antenna restrictions can impose significant constraints on the total information capacity.

\section{Extensions}
The key finding in this paper is a shape-independent bound on coupling strengths that we derive based on the domain-monotonicity property of the Green's function operator. 
This upper bound leads to two important discoveries. First, the sub-exponential decay in Eq.~(\ref{eq:decay3D}) identifies the slowest possible decay rate between any two domains in free space, and implies that three-dimensional domains have dramatically more channels available than their two-dimensional counterparts. Second, the ensuing bounds and scaling laws on the maximal number of usable communication channels and their maximal $N$-channel information capacity represent the ultimate limit that no domains can surpass. 
In this section, we briefly touch on other possible extensions of these results.

The bounding volume for the source and receiver domains can be any shape and size. We choose the concentric bounding volume in this article because of its analyticity and generality: its singular values are analytically tractable and the resulting bound is general enough to apply to any two domains that can be separated by a spherical surface.  In practice, if the sources and receivers are constricted to a domain smaller than the concentric bounding volume, one can sacrifice the analyticity by numerically computing the singular values of the largest possible domain for a tighter bound. Another analytical though less general bounding volume arises when the sources and receivers are known to be in the paraxial limit. Then, one can form the bounding volume as two rectangular cuboids whose singular values are known analytically in the paraxial limit~\cite{miller2000communicating}.  While we mainly focus on the concentric sphere--shell bounding volume in this work, future studies of alternative bounding volumes may reveal the dependence of the bound on the solid angles between the sources and receivers that otherwise cannot be captured by a concentric bounding volume.


Near-field information and power transfer have shown great promise in both wireless communication and fundamental science because of the abundant well-coupled channels in the form of electromagnetic evanescent waves~\cite{courjon1994near,want2011near, kim2015radiative,jawad2017opportunities}. This abundance emerges in our shape-independent bound in Eq.~(\ref{eq:decay3D}) where the optimal sub-exponential decay $\left(1 + d/R_{\text{min}}\right)^{-\sqrt{2q}}$ tends to unity when the separation distance $d$ goes to zero. Meanwhile, the maximal number of non-trivial communication channels diverges.
While this article mainly focuses on the application of our shape-independent bounds in the far field, it is also interesting to see how this formalism can regulate the maximal information and power transfer for different geometries in the near field.

The $n$-channel capacity bound proposed in this article may have ramifications on the optimal performance of antenna selections in massive multiple-input and multiple-output (MIMO) systems~\cite{molisch2004mimo,molisch2005capacity,gao2017massive,asaad2018massive}.
The technique of antenna selections mitigates the cost and complexity of MIMO systems by judiciously selecting only a fixed-size subset of antennas while maintaining a large total information capacity.  
How large the total information capacity can be among all the possible subsets is a question that falls under the umbrella of our $N$-channel capacity bound, which suggests the possibility to bound the capacity of any $N$-antenna subset by the capacity of the first $N$ optimal channels of the total antenna arrays. 

The presence of external scatterers can strongly affect the scattering amplitude of electromagnetic fields and the information content it carries. There are many shape-independent bounds proposed in this regard to bound the maximal power response of such external scatterers~\cite{sohl2007physical, yu2010fundamental,zhang2019scattering,presutti2020focusing,shim2020maximal,yang2018maximal,miller2016fundamental,miller2017limits,molesky2020global,gustafsson2020upper,kuang2020maximal,molesky2020hierarchical,kuang2020computational,miller2015shape,shim2019fundamental,molesky2019t,venkataram2020fundamental}, though there is still a need to understand their maximal information throughput. 
For example, to what degree could an external scatterer alter the sub-exponential decay rate predicted in this paper? What is the maximal number of non-trivial channels an external scatterer can help to establish and what are the maximal information capacities of those channels? 
Though a few bounds have been identified in certain physical scenarios~\cite{miller2007fundamental,miller2007fundamental2,molesky2021mathbb,shim2021fundamental},
those are still open questions that await for general answers.
Among various design techniques in search of better scatterer structures or antenna arrays, shape-independent bounds continue to offer a new lens to analyze the fundamental limits of information and power transfer in both fundamental physics and communication science.

\section{Acknowledgments}
Z.K. and O.D.M. were supported by the Army Research Office under grant number W911NF-19-1-0279. D.A.B.M was supported by the Air Force Office of Scientific Research under grant number FA9550-171-0002.

%

\end{document}


\title{Supplementary Material: Bounds on the Coupling Strengths of Communication Channels and Their Information Capacities: supplemental document}

\author{Zeyu Kuang}
\affiliation{Department of Applied Physics and Energy Sciences Institute, Yale University, New Haven, Connecticut 06511, USA}
\author{David A. B. Miller}
\affiliation{Ginzton Laboratory, Stanford University, 348 Via Pueblo Mall, Stanford, California 94305-4088,
USA}
\author{Owen D. Miller}
\affiliation{Department of Applied Physics and Energy Sciences Institute, Yale University, New Haven, Connecticut 06511, USA}

\date{\today}

\maketitle

\setcounter{tocdepth}{1}
\tableofcontents

\section{Singular values of the Green's function operator in the sphere--shell bounding volume}

In this section, starting from the full-electromagnetic wave equation, we define the dyadic Green's function and expand it with spherical vector waves. The spherical vector waves are the singular vectors of the dyadic Green's function operator in the sphere--shell bounding volume.  We present explicit expressions for these singular vectors and derive their corresponding singular values.  Results in this section supplement the arguments presented in Section 3A in the paper.

\subsection{Dyadic Green's function in spherical vector waves representation}
The dyadic Green's function  $\vG(\vrr, \vrr')$ in our paper is defined as the solution of the following wave equation under a point source excitation:
\begin{equation}
    \frac{1}{k^2}\curl{\curl{\vG(\vrr, \vrr')}} - \vG(\vrr, \vrr') = \vI\delta(\vrr - \vrr'),
    \label{eq:waveEq}
\end{equation}
where $\vI$ is the unit dyad and $k$ is the magnitude of the free-space wavevector.
Slightly different from the conventional definition by a factor of $k^2$, the wave equation in Eq. (\ref{eq:waveEq}) has the advantage of giving dimensionless singular values of the Green's function operator since $\vG(\vrr, \vrr')$ now has dimensions of reciprocal volume.
The solution of Eq. (\ref{eq:waveEq}) is commonly written as~\cite{tsang2004scattering}
\begin{equation}
    \vG(\vrr, \vrr') = (k^2 \vI + \nabla\nabla)\frac{e^{ik|\vrr-\vrr'|}}{4\pi|\vrr-\vrr'|}.
    \label{eq:GFrealspace}
\end{equation}
Considering the spherical symmetry of the sphere--shell bounding volume, we express the dyadic Green's function in Eq. (\ref{eq:GFrealspace}) in its spherical vector waves representation~\cite{tsang2004scattering}:
\begin{equation}
    \vG(\vrr, \vrr') = ik^3 \sum_{n=0}^\infty \sum_{m=-n}^n \sum_{j=1,2} \vv_{{\rm out,} nmj}(\vrr)\vv_{{\rm reg,} nmj}^*(\vrr'),
    \label{eq:GFsvw}
\end{equation}
where $\vv_{{\rm out,} nmj}(\vrr)$ and $\vv_{{\rm reg,} nmj}(\vrr')$ are the outgoing and regular spherical vector waves. Their explicit expressions are discussed in the following subsection. The index $n$ and $m$ are the two indices of the underlying spherical harmonics, and $j=1,2$ denotes the two possible polarizations of the transverse vector field. 

\subsection{Spherical vector waves as the singular vectors of the Green's function operator}
In this section, we give explicit expressions for the spherical vector waves, $\vv_{{\rm out,} nmj}(\vrr)$ and $\vv_{{\rm reg,} nmj}(\vrr')$. We also present a crucial orthogonal relationship for these spherical vector waves, which allows us to identify them as the singular vectors of the Green's function operator in the sphere--shell bounding volume.
The center of our sphere--shell bounding volume is chosen as the origin for the coordinates $\vrr$ and $\vrr'$. 
In spherical coordinates $(r,\theta,\phi)$, the spherical vector waves can be separated into a radial dependency of a spherical Hankel/Bessel function and an angular dependency of vector spherical harmonics. Their spatial distributions depend the polarization state $j = \{1, 2\}$, which we spell out separately:
\begin{align}
    \vv_{{\rm out}, nm1}(\vrr) &= \gamma_{n} h_n^{(1)}(kr)\vV^{(3)}_{nm}(\theta,\phi) \label{eq:out_vsw1} \\
    \vv_{{\rm out}, nm2}(\vrr) &= \gamma_{n}\left\{ n(n+1)\frac{h_n^{(1)}(kr)}{kr}\vV^{(1)}_{nm}(\theta,\phi) + \frac{\left[krh_n^{(1)}(kr)\right]'}{kr} \vV^{(2)}_{nm}(\theta,\phi) \right\} \label{eq:out_vsw2} \\
    \vv_{{\rm reg}, nm1}(\vrr) &= \gamma_{n} j_n(kr)\vV^{(3)}_{nm}(\theta,\phi) \label{eq:reg_vsw1} \\
    \vv_{{\rm reg}, nm2}(\vrr) &= \gamma_{n}\left\{ n(n+1)\frac{j_n(kr)}{kr}\vV^{(1)}_{nm}(\theta,\phi) + \frac{\left[krj_n(kr)\right]'}{kr} \vV^{(2)}_{nm}(\theta,\phi) \right\}, \label{eq:reg_vsw2}
\end{align}
where the prefactor $\gamma_{n}=1/\sqrt{n(n+1)}$. 
The three vector spherical harmonics are an extension of the scalar spherical harmonics: $\vV^{(1)}(\theta,\phi) = \hat{r} Y^m_n(\theta,\phi)$, $\vV^{(2)}(\theta,\phi) = r\nabla \left[Y^m_n(\theta,\phi)\right]$, and $\vV^{(3)}(\theta,\phi) = \nabla\cross\left[\hat{r} Y^m_n(\theta,\phi)\right]$, where $Y_n^m(\theta,\phi)=\sqrt{\frac{(2n+1)(n-m)!}{4\pi(n+m)!}} P^m_n(\cos\theta)e^{im\phi}$ are the scalar spherical harmonics defined by associated Legendre polynomials $P^m_n(x)$. The radial dependency of the outgoing spherical vector harmonics in Eqs. (\ref{eq:out_vsw1}, \ref{eq:out_vsw2}) are the spherical Hankel function of the first kind, $h^{(1)}_n(kr)$, with the domain of $r$ restricted to the region of the bounding shell. On the other hand, the regular  spherical vector harmonics $\vv_{{\rm reg,} nmj}(\vrr)$ are defined in the region of the bounding sphere. Because of this, their radial dependency follows the spherical Bessel function $j_n(kr)$. The regular spherical vector waves $\vv_{{\rm reg,} nmj}(\vrr)$ in Eqs. (\ref{eq:reg_vsw1}, \ref{eq:reg_vsw2}) take the same forms as their outgoing counterparts $\vv_{{\rm out,} nmj}(\vrr)$ but with every $h^{(1)}_n(kr)$ replaced by $j_n(kr)$.

The three vector spherical harmonics, $\vV_{nm}^{(1)}(\theta,\phi)$, $\vV_{nm}^{(2)}(\theta,\phi)$, and $\vV_{nm}^{(3)}(\theta,\phi)$, satisfy the following orthogonal property:
\begin{equation}
    \int_0^\pi d\theta \sin\theta\int_0^{2\pi}d\phi\vV_{nm}^{(\alpha)}(\theta,\phi)\cdot\vV_{n'm'}^{(\beta)*}(\theta,\phi) = z_{\alpha n}\delta_{\alpha\beta}\delta_{mm'}\delta_{nn'},
    \label{eq:Vorth}
\end{equation}
where the prefactor $z_{1n} = 1$ and $z_{2n} = z_{3n} = n(n+1).$
This orthogonality is crucial because it ensures that 1. different outgoing spherical vector waves $\vv_{{\rm out}, nmj}(\vrr)$ are orthogonal to each other in the outer bounding shell and 2. different regular spherical vector waves $\vv_{{\rm reg}, nmj}(\vrr)$ are orthogonal to each other in the inner bounding sphere. 
Considering these two orthogonal conditions and the fact that the Green's function operator in Eq.~(\ref{eq:GFsvw}) can be expanded as the sum of the outer products between $\vv_{{\rm out}, nmj}(\vrr)$ and $\vv_{{\rm reg}, nmj}(\vrr)$, we identify these two types of spherical vector waves as the left and right singular vectors of the Green's function operator in the sphere--shell bounding volume.


The discussion of the spherical vector wave representation in this subsection mostly follows the presentation in Ref. \cite[chapter~2.1]{tsang2004scattering}, though with different notations for the spherical vector waves and an extra factor of $k^2$ in the Green's function. We also adopt a more conventional definition of the scalar spherical harmonics as in Jackson \cite{jackson1999classical}. This leads to different prefactors in Eqs. (\ref{eq:out_vsw1} -- \ref{eq:Vorth}) compared to the ones in Ref. \cite{tsang2004scattering}.

\subsection{Singular values of the Green's function operator in the sphere--shell bounding volume}
Given the spherical wave expansion of the Green's function in Eq. (\ref{eq:GFsvw}), the singular values of the Green's function operator in the sphere--shell bounding volume can be identified as the products between the norms of the unnormalized singular vectors $\vv_{nmj}(\vrr)$ and $\text{Rg}\vv_{nmj}(\vrr)$ in their respective domains:
\begin{equation}
    |s_{nmj}^{\text{(sphere--shell)}}|^2 = k^6  \int_{V_{\text{sphere}}} \left|\vv_{{\rm reg,} nmj}(\vrr)\right|^2 d\vrr
    \int_{V_{\text{shell}}} \left|\vv_{{\rm out,} nmj}(\vrr)\right|^2 d\vrr .
    \label{eq:s_3Dintegral}
\end{equation}
The angular part of the integrals in Eq. (\ref{eq:s_3Dintegral}) can be computed by plugging in the explicit expressions of $\vv_{{\rm reg,} nmj}(\vrr)$ and $\vv_{{\rm out,} nmj}(\vrr)$ in Eqs. (\ref{eq:out_vsw1} -- \ref{eq:reg_vsw2}), which are simplified under the orthogonality relation of the vector spherical harmonics $\vV^{(i)}(\theta,\phi)$ in Eq. (\ref{eq:Vorth}).
The result is several remaining one-dimensional integrals in the radial direction:
\begin{align}
    |s^{\text{(sphere--shell)}}_{nm1}|^2 &= \int^{kR_{\text{sphere}}}_0 x^2 |j_n(x)|^2 dx \int^{kR_{\text{outer}}}_{kR_{\text{inner}}} x^2 |h^{(1)}_n(x)|^2 dx \label{eq:s_1Dintegral1}\\
    |s^{\text{(sphere--shell)}}_{nm2}|^2 &= \int^{kR_{\text{sphere}}}_0 \left\{n(n+1)|j_n(x)|^2 + \left[xj_n(x)\right]'^{\hspace{1pt}2} \right\} dx \nonumber \\ 
    & \qquad\qquad \times \int^{kR_{\text{outer}}}_{kR_{\text{inner}}} \left\{n(n+1)|h^{(1)}_n(x)|^2 + |xh^{(1)}_n(x)|'^{\hspace{1pt}2} \right\} dx, \label{eq:s_1Dintegral2}
\end{align}
where $R_{\text{sphere}}$ is the radius of the bounding sphere. The variables $R_{\text{inner}}$ and $R_{\text{outer}}$ are the inner and outer radii of the bounding shell, respectively.
The one-dimensional integrals in Eqs. (\ref{eq:s_1Dintegral1}, \ref{eq:s_1Dintegral2}) can be analytically integrated with the aid of indefinite integrals of the spherical Bessel functions in Ref. \cite{bloomfield2017indefinite}, after which we obtain explicit expressions of the singular values in the sphere--shell bounding volume:
\begin{align}
    |s^{\text{(sphere--shell)}}_{nm1}|^2 &= 
    \frac{\pi^2}{16}x^2\left[ J^2_{n+\frac{1}{2}}(x) - J_{n+\frac{3}{2}}(x)J_{n-\frac{1}{2}}(x)  \right]\eval_{x=0}^{x=kR_{\text{sphere}}} \nonumber \\ & \qquad\qquad\qquad\qquad  \times y^2\Re\left[ |H^{(1)}_{n+\frac{1}{2}}(x)|^2 - H^{(1)}_{n+\frac{3}{2}}(y)H^{(2)}_{n-\frac{1}{2}}(y)  \right]\eval_{y=kR_{\text{inner}}}^{y=kR_{\text{outer}}}  \label{eq:s_explicit1}
    \\
    |s^{\text{(sphere--shell)}}_{nm2}|^2 &= 
    \frac{\pi^2}{16} x^2\left\{ \frac{n+1}{2n+1} \left[ J^2_{n-\frac{1}{2}}(x) - J_{n+\frac{1}{2}}(x)J_{n-\frac{3}{2}}(x) \right] \right. 
    \nonumber \\ & \qquad\qquad\qquad\qquad \left. + \frac{n}{2n+1} \left[ J^2_{n+\frac{3}{2}}(x) - J_{n+\frac{5}{2}}(x)J_{n+\frac{1}{2}}(x) \right] \right\} \eval_{x=0}^{x=kR_{\text{sphere}}} \nonumber
    \\
    & \quad \times y^2\Re\left\{ \frac{n+1}{2n+1} \left[ |H^{(1)}_{n-\frac{1}{2}}(y)|^2 - H^{(1)}_{n+\frac{1}{2}}(y)H^{(2)}_{n-\frac{3}{2}}(y) \right] \right. 
    \nonumber \\ & \qquad\qquad\qquad\qquad \left. + \frac{n}{2n+1} \left[ |H^{(1)}_{n+\frac{3}{2}}(y)|^2 - H^{(1)}_{n+\frac{5}{2}}(y)H^{(2)}_{n+\frac{1}{2}}(y) \right] \right\} \eval_{y=kR_{\text{inner}}}^{y=kR_{\text{outer}}},
    \label{eq:s_explicit2}
\end{align}
where the functions $J_n(x)$ and $H^{(1)}_n(x)$ denote the Bessel function and the Hankel function of the first kind. Equations (\ref{eq:s_explicit1}, \ref{eq:s_explicit2}) are the explicit expressions of $|s^{\text{(sphere--shell)}}_{nmj}|^2$ we use in the paper to calculate the upper bounds of the coupling strengths between any two regions in the bounding volume.

\begin{figure}[b!]
\centering
\includegraphics[width=1\textwidth]{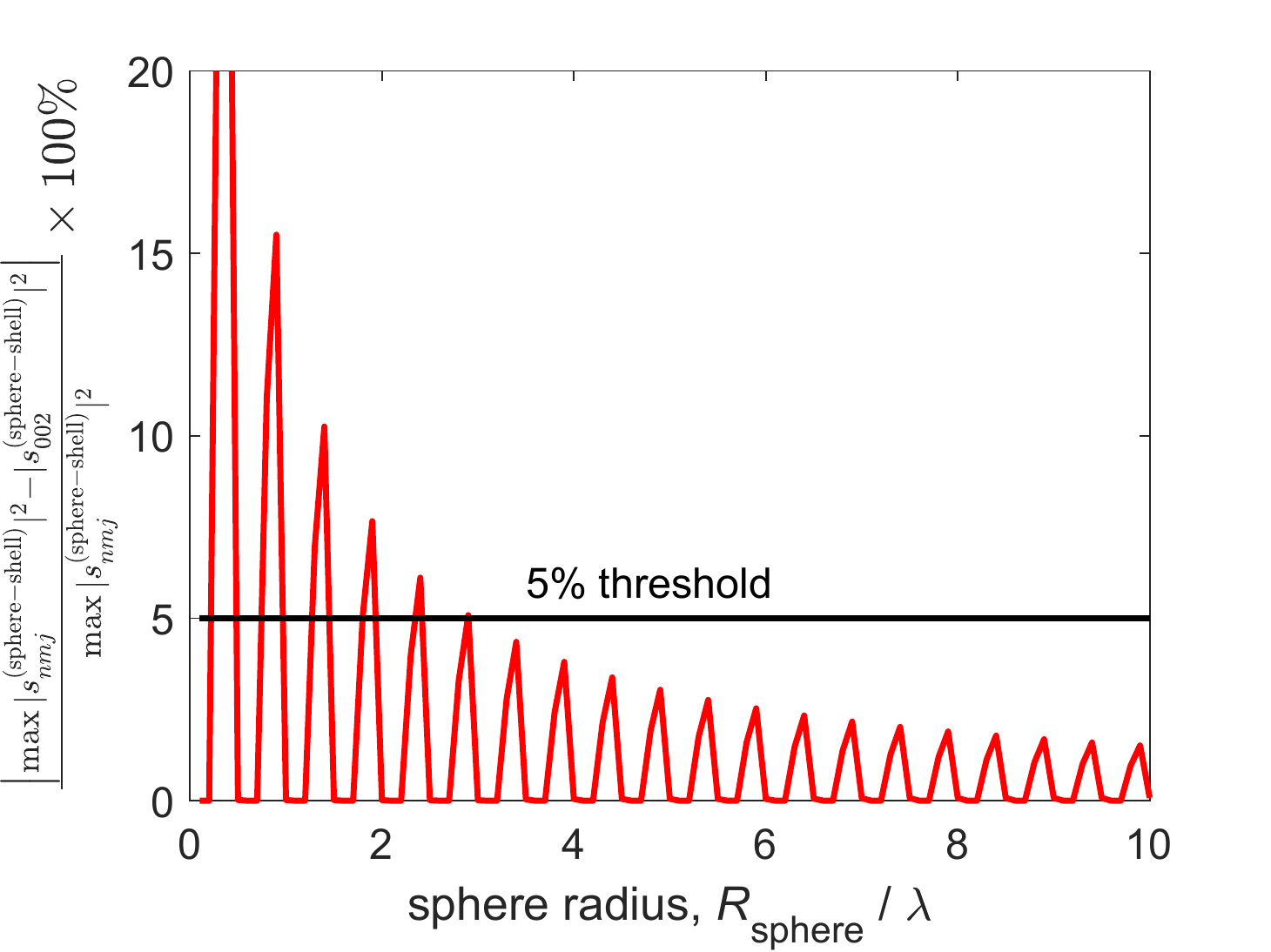}
\caption{Numerical evidence showing that the maximum channel strength, $\max{|s^{\text{(sphere--shell)}}_{nmj}|^2}$, is asymptotically attained by the first angular channel of the second polarization state, $|s^{\text{(sphere--shell)}}_{002}|^2$, in the limit of large sphere radius $R_s$ relative to the free-space wavelength $\lambda$.}
\label{fig:SM_Fig2}
\end{figure}

\section{Maximal channel strength in the limits of large bounding sphere and far-field bounding shells}
We observe that the maximal channel strength in the sphere--shell bounding domain is asymptotically attained by the first angular channel of the second polarization state in the limit of large bounding sphere:
\begin{equation}
    \max{|s^{\text{(sphere--shell)}}_{nmj}|^2} = |s^{\text{(sphere--shell)}}_{002}|^2,\quad \text{for\ } R_{\rm sphere} \gg \lambda, 
    \label{eq:sv_max_eq1}
\end{equation}
This is evidenced by Fig.~\ref{fig:SM_Fig2}, which shows that the relative difference between $\max{|s^{\text{(sphere--shell)}}_{nmj}|^2}$ and $|s^{\text{(sphere--shell)}}_{002}|^2$ is smaller than 5\% for a bounding sphere with radius $R_{\rm sphere}$ larger than three times the wavelength $\lambda$ and the relative difference asymptotically tends to zero as the radius becomes much larger than the wavelength. The separation distance $d$ between the two bounding domains and the maximal thickness $2R_r$ of the spherical shell is assumed to be $10\lambda$ and $\lambda$, respectively, though our result does not appear to be sensitive to these two parameters.

The channel strength  $|s^{\text{(sphere--shell)}}_{002}|^2$ has a simple analytical form in the limits of large bounding sphere and far-field bounding shell. To show this, we first consider the limit of far-field bounding shell, where $|s^{\text{(sphere--shell)}}_{002}|^2$, according to Eq.~(\ref{eq:s_1Dintegral2}), simplifies to
\begin{equation}
    |s^{\text{(sphere--shell)}}_{002}|^2 = 2kR_r\int_0^{kR_{\rm sphere}}x^2|j_{-1}(x)|^2 dx,\quad \text{for\ } d \gg \lambda, 
    \label{eq:sv_max_eq2}
\end{equation}
where we approximate the spherical Hankel function as $h^{(1)}_n(x) \approx \frac{1}{x}e^{ix}i^{-n-1}$ under the condition of $d \gg \lambda$. The integral in Eq.~(\ref{eq:sv_max_eq2}) can be analytically evaluated considering that $j_{-1}(x) = \cos(x)/x$. Its result, under the limit of large bounding sphere, further reduces to 
\begin{equation}
    |s^{\text{(sphere--shell)}}_{002}|^2 = k^2R_rR_{\rm sphere}, \ \text{for\ } R_{\rm sphere} \gg \lambda \text{\ and \ } d \gg \lambda.
    \label{eq:sv_max_eq3}
\end{equation}

Combining Eqs.~(\ref{eq:sv_max_eq1}) and~(\ref{eq:sv_max_eq3}), we derive an analytical expression of the maximal channel strength in the limits of large bounding sphere and far-field spherical shell:
\begin{equation}
    \max{|s^{\text{(sphere--shell)}}_{nmj}|^2} = k^2R_rR_{\rm sphere},\ \text{for\ } R_{\rm sphere} \gg \lambda \text{\ and \ } d \gg \lambda,
    \label{eq:sv_max_eq4}
\end{equation}
which, we observe, scales linearly with the maximal radii of both the source and receiver domains.

\section{A lower bound on the sum rule}
The sum rule $S=\sum_{nmj}|s_{nmj}|^2$ is conserved under a unitary transformation from the communication channel basis to the delta-function basis in real space. Conveniently, we express $S$ as a double integral of the Frobenius norm of the dyadic Green's function over both the source and receiver volumes:
\begin{equation}
     S = \int_{V_s}\int_{V_r} ||\vG(\vrr,\vrr')||_F^2 d\vrr d\vrr'.
\end{equation}
The Frobenius norm of the dyadic Green's function reads~\cite{miller2016fundamental}
\begin{equation}
    ||\vG(\vrr,\vrr')||_F^2 = \frac{k^6}{8\pi^2}\left[ \frac{1}{(k|\vrr-\vrr'|)^2} + \frac{1}{(k|\vrr-\vrr'|)^4} + \frac{3}{(k|\vrr-\vrr'|)^6}\right]
\end{equation}
which monotonically decays with respect to the separation distance, $|\vrr-\vrr'|$, between two points. This monotonic decay allows us to lower bound the sum rule by relaxing the separation distance to the largest possible separation distance, $\max{|\vrr-\vrr'|} = d+2R_s+2R_r$, between the source and receiver volumes:
\begin{equation}
    S \geq \frac{k^4V_sV_r}{8\pi^2(d+2R_s+2R_r)^2} + \mathcal{O}\left(\left[k(d+2R_s+2R_r)\right]^{-4}\right). \label{eq:sum}
\end{equation}
The variables $R_s$ and $R_r$ denote the maximal radii of the source and receiver domains.
For conciseness, we assume the furthest separated points are in the far field, i.e. $k(d+2R_s+2R_r) \gg 1$, so  that only the leading term in Eq. (\ref{eq:sum}) remains. This, of course, can be easily generalized by explicitly including two other higher-order terms with a slightly more complicated expression. 

\section{An upper bound on the relative coupling strengths in the large-channel limit}
In this section, we derive large-channel asympotes of the coupling strengths in the sphere--shell bounding volume. 
Two differently polarized communication channels exhibit slightly different asymptotes, though both can be bounded above by a single expression. 
Together with the lower bound on the total sum rule, we derive an upper bound on the relative coupling strength for both polarizations. This section provides a theoretical basis for the bound we present at the end of Section 3A in the paper.

The singular values $|s^{\text{(sphere--shell)}}_{nmj}|^2$ have simple analytical expressions in the large-channel limit when the index $n\rightarrow\infty$. They can be derived by substituting the large-$n$ asymptotes of the spherical Bessel and Hankel functions, $j_n(x) \sim \frac{1}{\sqrt{(4n+2)x}} \left(\frac{ex}{2n+1}\right)^{n+1/2}$ and $h^{(1)}_n(x) \sim \frac{-2i}{\sqrt{(4n+2)x}}\left(\frac{ex}{2n+1}\right)^{-n-1/2}$, into Eq. (\ref{eq:s_1Dintegral1}, \ref{eq:s_1Dintegral2}):
\begin{align}
    |s^{\text{(sphere--shell)}}_{nm1}|^2 &= \left( \frac{kR_{\text{sphere}}}{2n} \right)^4 \left( \frac{R_{\text{sphere}}}{R_{\text{inner}}} \right)^{2n-1} \quad \text{as\ } n\rightarrow\infty, \label{eq:s_asym1}
    \\
    |s^{\text{(sphere--shell)}}_{nm2}|^2 &= \frac{1}{4} \left( \frac{R_{\text{sphere}}}{R_{\text{inner}}} \right)^{2n+1} \quad \text{as\ } n\rightarrow\infty. \label{eq:s_asym2}
\end{align}
While both polarizations decay exponentially as a function of $n$, the first polarization channel is always smaller than the second one in the large $n$ limit due to the additional decay of the factor of $1/n^4$. The value of the second polarization thus serves as an upper bound for both:
\begin{equation}
    |s^{\text{(sphere--shell)}}_{nmj}|^2 \leq \frac{1}{4} \left( \frac{R_{\text{sphere}}}{R_{\text{inner}}} \right)^{2n+1} \quad \text{as\ } n\rightarrow\infty.
    \label{eq:s_asym_bd}
\end{equation}
This is an upper bound for the coupling strengths of both polarizations in a sphere--shell bounding volume in the large-channel limit. The bound only depends on the ratio between the radius of the bounding sphere, $R_{\text{sphere}}$, and the inner radius of the bounding shell, $R_{\text{inner}}$ --- the smaller the ratio, the faster the decay.

For any two domains that can be separated by a spherical surface, there are two possible sphere--shell bounding volumes: one that centers around the source region and one that centers around the receiver region. To obtain a tighter upper bound, we choose the one that centers around the smaller domain because it has the smaller ratio between $R_{\text{sphere}}$ and $R_{\text{inner}}$. Considering this and the fact that the number of channels with $n$-index less or equal to $n$ is $q=2(n+1)^2$, Eq. (\ref{eq:s_asym_bd}) can be written as
\begin{equation}
    |s^{\text{(sphere--shell)}}_{q}|^2 \leq \frac{1}{4} \left( 1 +  \frac{R_{\text{min}}}{d} \right)^{-\sqrt{2q}-1} \quad \text{as\ } q\rightarrow\infty,
    \label{eq:s_asym_bd_l}
\end{equation}
where $R_{\text{min}} = \text\{R_s, R_r\}$ is the smaller of the radius of the source domain $R_s$ and the radius of the receiver domain $R_r$, and $d$ is the distance between the two domains. This equation shows the coupling strengths between two regions always decay \textit{sub}-exponentially with the total channel index, $q$, in the large-channel limit.

Lastly, we invoke the domain-monotonicity theorem discussed in Section IIA of the paper which implies that the coupling strengths $|s_q|^2$ between any two domains have to be smaller than their counterparts in a sphere--shell bounding volume:
\begin{equation}
    |s_{q}|^2 \leq |s_q^{\text{(sphere--shell)}}|^2,\quad  \text{for}\ q = 1, 2, ...
    \label{eq:sl_max}
\end{equation}
Combining this with the upper bound of $|s^{\text{(sphere--shell)}}_{q}|^2$ in Eq. (\ref{eq:s_asym_bd_l}) and the lower bound of the sum rule $S$ in Eq. (\ref{eq:sum}), we derive an upper bound on the relative channel strengths in the large-channel limit:
\begin{equation}
    \frac{|s_{q}|^2}{S} \leq \frac{2\pi^2(d+2R_s+2R_r)^2}{k^4V_sV_r(1+d/R_{\text{min}})^{\sqrt{2q}+1}}, \quad \text{as}\ q \rightarrow \infty.
    \label{eq:decay3D}
\end{equation}
This suggests that the relative coupling strength between any two domains decay at least sub-exponentially in the large-channel limit. Equation (\ref{eq:decay3D}) is a key result presented in our paper and we hereby provide a derivation in this section.

\begin{figure*}[t]
\centering
\includegraphics[width=1\textwidth]{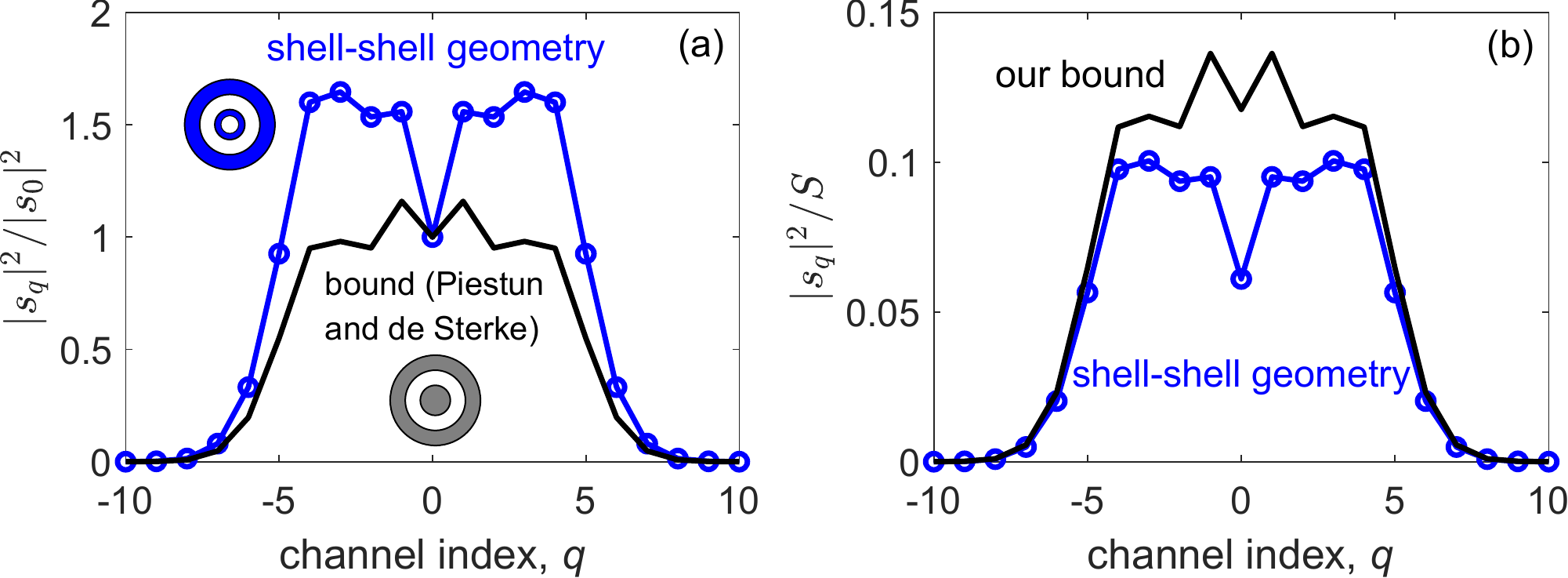}
\caption{\label{fig:sv_bounds} A shell--shell structure (blue-colored inset) that violates the propsed bound by Piestun and de Sterke in Ref. \cite{piestun2009fundamental}. (b) The relative coupling strength $|s_l|^2 / |s_0|^2$ of the shell--shell geometry, normalized by the 0-th order coupling strength, violates the proposed bound by Piestun and de Sterke \cite{piestun2009fundamental} which are their counterparts in the cylinder--shell bounding volume (grey-colored inset). (c) The relative coupling strength $|s_l|^2 / S$ from the same shell--shell structure, normalized by the sum rule, is correctly bounded in our approach. }
\label{fig:SM_Fig1}
\end{figure*}

\section{Comparison with the results of Piestun and de Sterke}
Piestun and De Sterke~\cite{piestun2009fundamental} have analyzed concentric cylindrical objects to obtain a first approximate analysis of the numbers of well-coupled communications modes in two dimensions. 
The reason it is an approximation, not a exact bound, is because of two assumptions they made. First, they assume, for all geometries, the channel strengths $|s_q|^2$ are constant up to a certain channel index, after which the channel strengths fall off rapidly. Second, they assume, for all geometries, their relative channel strengths $|s_q|^2 / |s_0|^2$ are upper bounded by their counterparts in the cylinder--shell bounding volume, $|s^{\text{(cylinder--shell)}}_q|^2 / |s^{\text{(cylinder--shell)}}_0|^2$. Under these two assumptions, they derive an upper bound on the number of well-coupled channels, $N$, in a cylinder--shell bounding volume:
\begin{equation}
    N \approx \sum_{q=-\infty}^{\infty} \frac{|s_q|^2}{|s_0|^2} \leq \sum_{q=-\infty}^{\infty} \frac{|s^{\text{(cylinder--shell)}}_q|^2}{|s^{\text{(cylinder--shell)}}_0|^2}.
    \label{eq:piestun}
\end{equation}
This expression is meaningful for (circular) cylinders, but is not a ``fundamental limit'' for any shape. In order for the inequality in Eq.~(\ref{eq:piestun}) to be valid, one would need the denominator on the right, the first singular value of the cylinder, to be less than or equal to the denominator on the left, the first singular value of any arbitrary domain. Yet this inequality is not valid in general.
One geometry that violates the singular-value inequality, and as a result also the bound of Eq.~(\ref{eq:piestun}), is a ``shell--shell'' geometry that consists of two concentric cylindrical shells as shown in the blue-colored inset of Fig. \ref{fig:SM_Fig1}(a). Its normalized channel strengths, $|s_q|^2 / |s_0|^2$, are plotted as the circled blue line for channel index $q$ from -10 to 10. We assume the inner cylindrical shell has inner and outer radii of $0.3\lambda$ and $\lambda$, and the outer cylindrical shell has inner and outer radii of $10\lambda$ and $11\lambda$. Nearly all the normalized channel strengths, $|s_q|^2 / |s_0|^2$, of the shell--shell structure surpass the supposed upper bound (black solid line) given by the normalized channel strengths, $|s^{\text{(cylinder--shell)}}_q|^2 / |s^{\text{(cylinder--shell)}}_0|^2$, of the cylinder--shell bounding volume. 
Consequently, the maximal number of channels $N = \sum_{l=-\infty}^{l=\infty} |s^{\text{(cylinder--shell)}}_l|^2 / |s^{\text{(cylinder--shell)}}_0|^2 = 10.7$ predicted from Eq. (\ref{eq:piestun}) is a number smaller than the one obtained with the actual shell--shell structure $N = \sum_{l=-\infty}^{l=\infty} |s_l|^2 / |s_0|^2 = 16.4.$ 

On the other hand, the approach presented in our paper can correctly bound the response from the same shell--shell geometry as shown in Fig.~\ref{fig:SM_Fig1}(b). Our bound is in fact a rigorous upper bound to all geometries  because we do not make any prior assumptions. We do not assume a step-like distribution of the channel strengths (as in~Ref.\cite{piestun2009fundamental}). We also employ a sum-rule normalization where the relative channel strengths can always be bounded above. 

%